\documentclass{article}
\usepackage[T1]{fontenc} 
\usepackage[utf8]{inputenc} 
\usepackage{ismir,amsmath,cite,url}
\usepackage{graphicx}
\usepackage{color}
\usepackage[dvipsnames]{xcolor}
\usepackage{units}
\usepackage{gensymb}
\usepackage{paralist}
\usepackage{amssymb}
\usepackage{microtype}

\newcommand{\jiawen}[1]{#1} 

\newcommand{\ashis}[1]{#1}
\newcommand{\amy}[1]{#1}

\title{Score-informed Networks for Music Performance Assessment}

\multauthor
{Jiawen Huang\thanks{These authors contributed equally to this work.} \hspace{1cm} Yun-Ning Hung\footnotemark[1] \hspace{1cm} Ashis Pati} { \bfseries{Siddharth Gururani \hspace{1cm} Alexander Lerch}\\
  Center for Music Technology, Georgia Institute of Technology, USA\\
{\tt\small \{jhuang448,amyhung,ashis.pati,siddgururani,alexander.lerch\}@gatech.edu}
}
\def\authorname{Jiawen Huang, Yun-Ning Hung, Ashis Pati, Siddharth Gururani, Alexander Lerch}

\begin{document}

\maketitle
\begin{abstract}
  The assessment of music performances in most cases takes into account the underlying musical score being performed. While there have been several automatic approaches for objective music performance assessment (MPA) based on extracted features from both the performance audio and the score, deep neural network-based methods incorporating score information into MPA models have not yet been investigated.
  In this paper, we introduce three different models capable of score-informed performance assessment. These are 
  \begin{inparaenum}[(i)]
    \item a convolutional neural network that utilizes a simple time-series input comprising of aligned pitch contours and score,
    \item a joint embedding model which learns a joint latent space for pitch contours and scores, and
    \item a distance matrix-based convolutional neural network which utilizes patterns in the distance matrix between pitch contours and musical score to predict assessment ratings.
  \end{inparaenum}
  Our results provide insights into the suitability of different architectures and input representations and demonstrate the benefits of score-informed models as compared to score-independent models.
\end{abstract}

\section{Introduction}
\label{sec:intro}

  A performance is a sonic rendition of a written musical score (in the case of Western classical music). The characteristics of a music performance play a major role in how listeners perceive music, even if performances are based on the same underlying score \cite{lerch19mpa}. To perform a musical piece, the performer must first parse the score, interpret or  modify the musical information, and  utilize complex motor skills to render the piece on their instrument \cite{palmer1997music}.

  From the perspective of the performer, mastery over the art of music performance is often a journey spanning several years of instruction and practice. A major factor in learning and improving one's skill as a performer is to analyze and obtain feedback regarding the performance. Due to the complex nature of music performance, students require regular feedback from trained professionals. Teachers are expected to grade or rate students based on various performance criteria such as note accuracy or musicality. These criteria are often ill-defined and subject to interpretation, thus making objective and consistent music performance assessment (MPA) rather difficult \cite{wesolowski2016examining,thompson2003evaluating}. Regardless, this subjective manner of MPA is still used, e.g., in school systems where ensemble members are selected based on instructors' assessments of student auditions. 

  Wu et al.\ discussed the notion of objective descriptors (features) which are potentially useful for automatic MPA \cite{wu16towards}. Such features are computed by applying signal processing methods to recorded performances and are used to model teachers' assessments of the performances using machine learning. With the rise of deep learning, neural networks were found to outperform the classical pipeline of feature extraction followed by regression \cite{pati2018assessment}. However, one issue with these approaches is that they ignore the score that the students are meant to play. We will refer to such approaches as \textit{score-independent}. The idea of incorporating score-based features utilizing audio to score alignment was explored, e.g., by Vidwans et al.\ \cite{vidwans17objective}. Further analysis of hand-crafted features for MPA showed the relative importance of score-based features over score-independent ones \cite{gururani18analysis}. Therefore, the design of deep architectures that incorporate score information is an obvious and overdue extension of previous approaches.

  The goal of this paper is to explore different methods to incorporate this score information. Our hypothesis is that including score information will lead to improved performance of deep networks in the objective MPA task. To this end, we present three architectures which combine score and audio features to make a \textit{score-informed} assessment of a music performance. First, we concatenate aligned pitch contours and scores into a 2-dimensional time-series feature representation that is fed to a convolutional neural network (CNN). Second, we propose a joint embedding model for aligned score and pitch contours. The assessment ratings are predicted using the cosine similarity between the score and performance embeddings. Third, we utilize the distance matrix, a mid-level representation combining both the score and pitch contour, as the input to a deep CNN trained to predict the teachers' assessments. Finally, using a fairly large scale dataset of middle school and high school student auditions, we perform an in-depth evaluation comparing these proposed architectures against each other and with a score-independent baseline approach for MPA . 
  

\section{Related Work}
\label{sec:relwork}
  MPA deals with the task of assessing music performances based on audio recordings. Progress in MPA is roughly categorized into feature design-based approaches \cite{knight2011potential,nakano2006automatic,wu16towards,romani2015real,gururani18analysis} and feature learning-based approaches \cite{wu2018learned,pati2018assessment,han2014hierarchical}. Feature design-based methods rely on signal processing techniques to either extract standard spectral and temporal features \cite{knight2011potential}, or use expert knowledge to extract perceptually motivated features capable of characterizing music performances \cite{nakano2006automatic,romani2015real}. The extracted features are then fed into simple machine learning classifiers to train models which predict different performance assessment ratings. Feature learning-based approaches, on the other hand, stem from the argument that important features for modeling performance assessments are not trivial and cannot be easily described. Hence, they rely on using mid-level representations (such as pitch contours or mel-spectrograms) as input to sophisticated machine learning models such as sparse coding \cite{han2014hierarchical,wu2018learned} and neural networks \cite{pati2018assessment}. 

  Most performances of Western music, however, are based on written musical scores. Hence, performances are also assessed based on their perceived deviations from the underlying score. 
  There has been some prior research on incorporating the score information into the assessment modeling process. Most of the these approaches rely on computing descriptive features using some notion of \textit{distance} between the score representation and the performance representation \cite{mayor2009performance,devaney_study_2012,molina2013fundamental,falcao_dataset_2019,huang_automatic_2019}. The most common approach has been to first use an alignment algorithm, e.g., Dynamic Time Warping (DTW) \cite{sakoe_dynamic_1978}, to temporally align the performance recording with the score and then compute descriptive features which characterize the deviations of the performance from the score \cite{vidwans17objective,bozkurt2017voice}. However, to the best of our knowledge, incorporating score information directly into neural network-based models for MPA has not been investigated before. 

  Score-informed approaches have helped improve results for both related performance analysis tasks and other music information retrieval tasks. Most of these methods have also relied on an alignment between the audio recording and the score as the primary tool for incorporating score information. Aligning audio recordings with scores has been useful for several tasks such as detecting expressive features in music performances \cite{li2015analysis}, identifying missing notes and errors in piano performances \cite{ewert2016score}, and segmenting syllables in vocal performances\cite{pons2017score}. Scores have also been used to generate soft labels and/or artificial training data for tasks such as source separation \cite{miron2017monaural,ewert2017structured}.

\section{Methods}
\label{sec:method}
  We propose and compare three different approaches to incorporate the score information with audio features for MPA.\footnote{The code is available at: https://github.com/biboamy/FBA-Fall19} 
  The score information is represented as the MIDI pitch sequence (in ticks) obtained from the sheet music of the score to be performed. Henceforth, the MIDI pitch sequence will be referred to as the \textit{score}. The student's \textit{performance} is represented by the pitch contour of the audio. We use pitch contour since it captures both pitch and rhythmic information. Musical dynamics and timbre are ignored in this study; while dynamics are an important expressive tool for the performer \cite{lerch19mpa}, the score usually lacks specificity in dynamics instructions and cannot serve as the same absolute reference as for pitch and rhythm. 

  \subsection{Score-Informed Network (SIConvNet)}
  \label{sec:si_network}
    \begin{figure}[t]
      \centering
      \includegraphics[width =1\columnwidth]{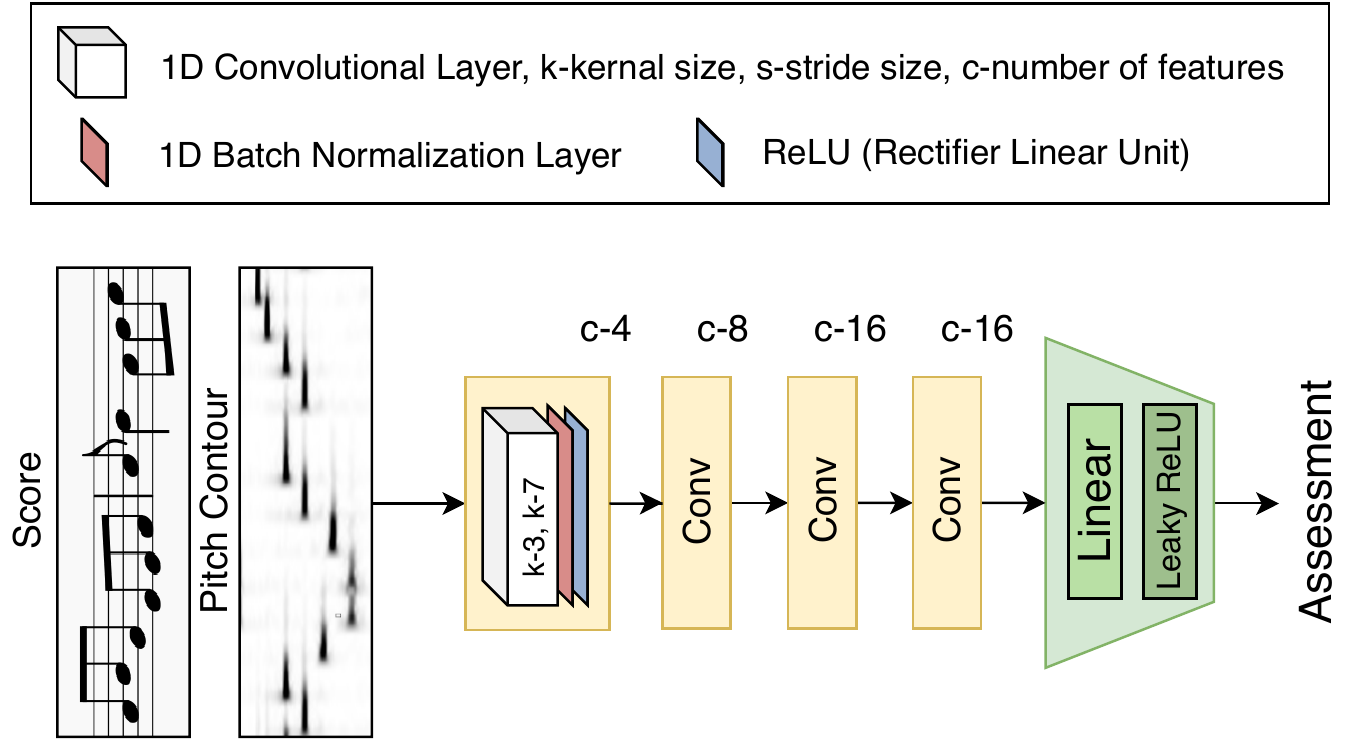}
      \caption{Schematic for the SIConvNet. The aligned \textit{score} and \textit{pitch contour} are stacked together and fed into a 4-layer CNN to directly predict the assessment ratings.}
      \label{fig:si_model}
    \end{figure}

    The first approach that we use is probably the most straightforward way of incorporating score information into the assessment model. A simple CNN is used that relies on both the score and performance as the input and directly predicts the assessment ratings.

    \subsubsection{Input Representation}
    \label{sec:input_rep}
      The input representation for this approach is constructed by simply stacking an aligned pitch contour and score pair to create a $N\times 2$ matrix, where $N$ is the sequence length of the pitch contour. \ashis{The two channels correspond to the pitch contour and score, respectively.}
      
      In order to obtain this representation, we first consider a pitch contour snippet of length $N$ \amy{(sequence of logarithmic frequencies)}. Then, we find the corresponding part of the score using a DTW-based alignment process. The obtained score snippet \amy{(sequence of MIDI note numbers)} is then resampled to have the same length $N$ as the pitch contour.
      
    \subsubsection{Model Architecture}
     A schematic of the model architecture is shown in \figref{fig:si_model}. We use a simple 4-layer CNN based on the architecture proposed by Pati et al.\ \cite{pati2018assessment} and append a single linear layer which predicts the assessment. Each convolutional stack consists of a 1-D convolution followed by a 1-D batch normalization layer \cite{ioffe2015batch} and ReLU non-linearity. The linear layer at the end comprises of a dense layer followed by Leaky ReLU non-linearity. 

  \subsection{Joint Embedding Network (JointEmbedNet)}
  \label{sec:je_network}

    \begin{figure}[t]
      \centering
      \includegraphics[width =1\columnwidth]{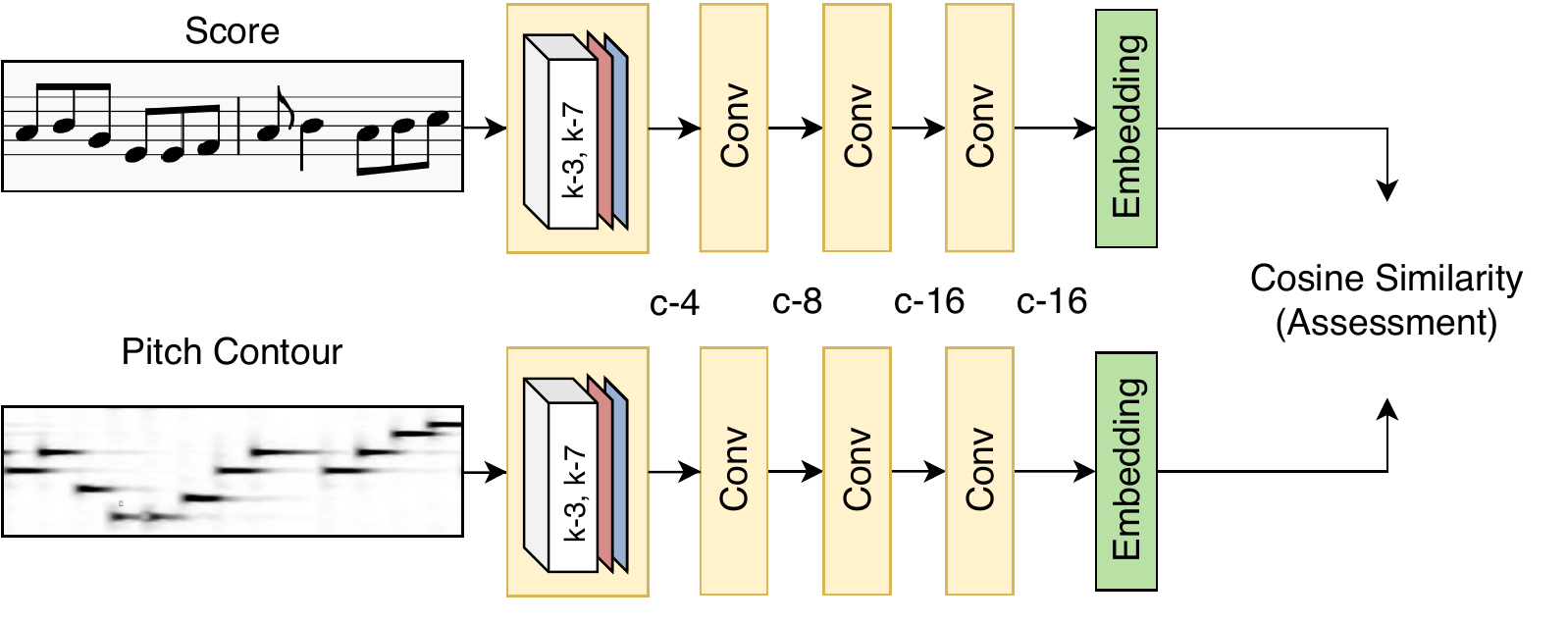}
      \caption{Schematic of the JointEmbedNet architecture.}
      \label{fig: je_model}
    \end{figure}

    \ashis{The second approach is based on the assumption that performances are rated based on some sort of perceived distance between the performance and the underlying score being performed.}
    Consequently, we use two separate encoder networks to project the score and the pitch contour to a joint latent space and then use the similarity between the embeddings to predict the assessment ratings. 



  \subsubsection{Input Representation}
    This approach uses the same input representation as SIConvNet (see \secref{sec:input_rep}). However, instead of stacking the aligned pitch contour and the score, the individual $N \times 1$ sequences are fed separately to the two encoders. 

  \subsubsection{Model Architecture}
    This network (see \figref{fig: je_model}) uses two 1-D convolutional encoders having the same architecture as SIConvNet. Each encoder has 4 convolutional blocks to extract high level feature embeddings. The performance encoder is expected to extract relevant features pertaining to the performance from the pitch contour. On the other hand, the score encoder is expected to extract the important features from the score. Assuming that the assessment rating for the performance is high if these two embeddings are similar, we use the cosine similarity $\mathrm{cos}(E_\mathrm{score}, E_\mathrm{performance})$ between the two embeddings to obtain the predicted assessment rating. $E_\mathrm{score}$ and $E_\mathrm{performance}$ are the embeddings obtained from the score and performance encoders, respectively. \ashis{If the two embeddings are similar}, the cosine similarity is close to one, and the model \jiawen{will} predict a higher rating.

  \subsection{Distance Matrix Network (DistMatNet)}
  \label{sec:dm_network}

  The final approach uses a distance matrix between the pitch contour and the score as the input to the network. Given the information from both the \ashis{pitch contour and the score}, the task of performance assessment might be interpreted as finding a \textit{performance distance} between them. Thus, the choice of the distance matrix as the input representation allows the model to learn from the pitch differences. A Residual CNN \cite{He2016DeepRL} architecture is chosen for the network. 


  \begin{figure}[t]
    \centering
    \includegraphics[width =1\columnwidth]{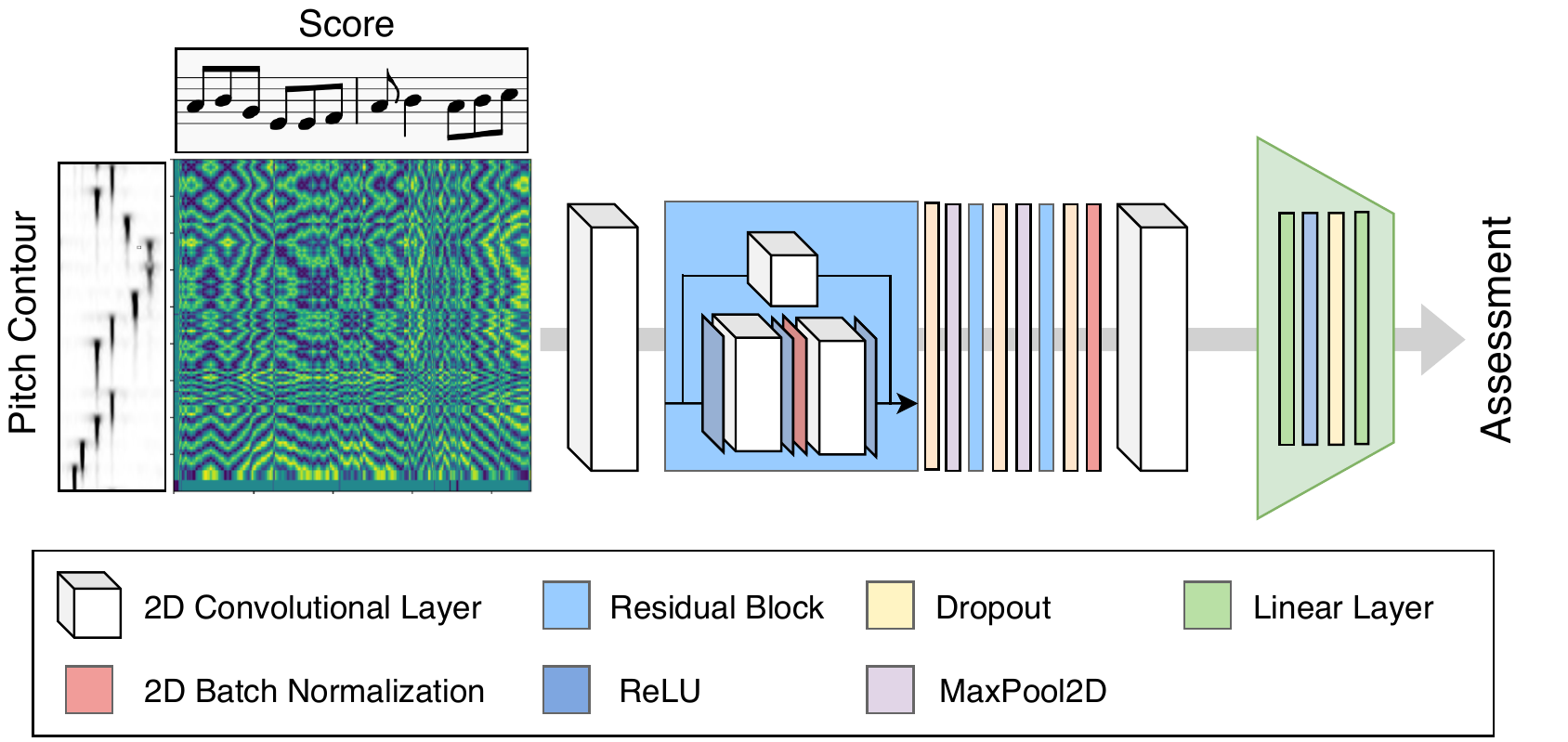}
    \caption{Schematic of the DistMatNet architecture.}
    \label{fig: dm_model}
  \end{figure}

  \subsubsection{Input Representation}

    The distance matrix elements are the pair-wise wrapped distances between the pitch contour and the MIDI pitch sequence. The octave-independent wrapped-distance is used to compensate for possible octave errors made by the pitch tracker. To ensure a uniform input size to the network, the matrix is resampled to a square shape of a fixed size. Thus, a performance with constant tempo would result in an aligned path located on the diagonal. Unlike the previous two methods where the input pitch contour and the score are aligned using DTW, the distance matrix input avoids any error propagation caused by alignment errors.
    The choice of this input representation stems from the success of distance matrices (or self-similarity matrices) in other areas of MIR such as structural segmentation \cite{grill2015music,cohen2017music} and music generation \cite{wei2019generating}.


  \subsubsection{Model Architecture}
    The model architecture is shown in \figref{fig: dm_model}. It is composed of 3 residual blocks. Each residual block has 2 convolutional layers. Dropout and max-pooling are added between each residual block. A classifier with two linear layers (128 features) with one ReLU and dropout layer in between is used after the residual network to perform regression prediction. \amy{We use (3,3) kernal size and 4 feature maps for all convolutional layers, 0.2 dropout rate, and a (3,3) kernal size for all max-pooling layers.}

\section{Experiments}
\label{sec:exp}

\subsection{Dataset}

The dataset we use to evaluate our methods is a subset of a large dataset of middle school and high school student performances. These are recorded for the Florida All State Auditions, which are separated into three bands: 
\begin{inparaenum}[(i)]
  \item middle school band,
  \item concert band, and 
  \item symphonic band.
\end{inparaenum}
The recordings contain auditions spanning 6 years (from 2013 to 2018), and feature several \ashis{monophonic} pitched and percussion instruments. Each student performs rehearsed scores, scales and a sight reading exercise. For the purpose of this study we limit our experiments to \ashis{the \textit{technical etude}} for middle school and symphonic band auditions. We choose \textit{Alto Saxophone}, \textit{Bb Clarinet} and \textit{Flute} performances due to these being the most popular across all pitched instruments. \tabref{tab: data_dist} shows the distribution of data across different instruments. \ashis{The average duration of each performance is \unit[30]{s} for middle school and \unit[50]{s} for symphonic band students.}
\ashis{The dataset also includes the musical scores that the students are supposed to perform for each exercise}. \jiawen{The average length (in notes) of the musical scores are 136 for middle school and 292 for symphonic band.} Note that \ashis{these scores} are the same across all students performing the same instrument in the same year but vary across years and instruments. 

The dataset also contains expert assessments for each exercise of a student performance. \ashis{Each performance is rated by one expert along} $4$ criteria \ashis{defined by the Florida Bandmasters' Association (FBA)}: 
\begin{inparaenum}[(i)]
  \item musicality,
  \item note accuracy, 
  \item rhythmic accuracy, and
  \item tone quality.
\end{inparaenum}
\ashis{All ratings are on a point-based scale and are normalized to range between $0$ to $1$ by dividing by the maximum.}
Since we focus on pitch contours as the primary audio feature, tone quality is excluded from this study. 

\begin{table}
  \begin{tabular}{l|c|c}
                 & \multicolumn{1}{l|}{Middle School} & \multicolumn{1}{l}{Symphonic Band} \\ \hline
  Alto Saxophone & {\color[HTML]{000000} 696}         & {\color[HTML]{000000} 641}         \\ \hline
  Bb Clarinet    & {\color[HTML]{000000} 925}         & {\color[HTML]{000000} 1156}        \\ \hline
  Flute          & {\color[HTML]{000000} 989}         & {\color[HTML]{000000} 1196}       
  \end{tabular}
  \caption{Number of performances for the different instruments per band.}
  \label{tab: data_dist}
  \end{table}

\subsubsection{Data pre-processing}

The pitch contours are extracted using the pYIN algorithm \cite{mauch2014pyin} with a block size and hop size of $1024$ and $256$ samples, respectively. The audio sampling rate is \unit[44100]{Hz}. The extracted frequencies are converted from Hz to MIDI pitch \jiawen{(unlike the MIDI pitches from the musical score, these can be floating point numbers)}. Both the resulting pitch contour and musical score are normalized by dividing by $127$. 
Finally, for the purpose of model training and evaluation, we divide our dataset into three randomly sampled subsets: training, validation, and testing. We use a ratio of $8\colon1\colon1$ for splitting the dataset.

We use \textit{random-chunking} as a data augmentation tool when training SIConvNet and JointEmbedNet since it has shown to be useful in improving model performance \cite{pati2018assessment}. First, the pitch contour is chunked into snippets of length $N$ by randomly selecting the starting position. The corresponding aligned and length-adjusted score snippet is obtained using the method described in \secref{sec:input_rep}.
We assume the chunked segment has the same assessment score as the whole recording. 
We do not perform chunking on our distance matrix since the matrix has already been resampled into a smaller resolution. Instead, we discuss how varying the resampling size could effect the performance in one of the experiments. 



\subsection{Experimental Setup}

We present three experiments to evaluate our proposed methods. 
First, we compare the overall performance of the proposed architectures against a score-independent baseline system PCConvNet \cite{pati2018assessment} which uses only the randomly-chunked pitch contour as input. This experiment also gives us an indication of the effectiveness of each of the proposed methods. Second, we look at the sensitivity of the SIConvNet and JointEmbedNet to the chunk size $N$. Finally, we investigate the effect of varying the resolution of the input distance matrix for the DistMatNet model. The latter two experiments were aimed at understanding the effects of the different hyper-parameters used while constructing the input data for each model. These helped us arrive at the best parameters for each approach.

\begin{table}[]
    \centering
    \begin{tabular}{c|c|c}
      \phantom{aa}SIConvNet\phantom{aa} & JointEmbedNet & \phantom{aa}DistMatNet\phantom{aa}\\ \hline
      3,089&6,144&63,417
    \end{tabular}
    \caption{\amy{Number of parameters for each model.}}
    \label{tab:parameters}
\end{table}

\ashis{The number of trainable parameters for each method is shown in Table~\ref{tab:parameters}. DistMatNet has a higher number of parameters because it uses a higher-dimensional input with a deeper architecture to capture high level information \cite{He2016DeepRL}.}

For each method, we trained separate models to predict each assessment criterion. Moreover, to measure the variation of each model, we trained each model on 10 different random seeds. We used $M_i$ to represent the model training on different random seed where $i = 0 \ldots 9$. A boxplot with median 
and variation of each $M_i$ is shown to demonstrate the results. \amy{All the models are trained based on the mean squared error between estimated and ground truth ratings.}  All the models are trained with a stochastic gradient descent optimizer with a 0.05 learning rate. We apply early stopping if the validation loss does not improve for 100 epochs. The performance of all models is measured using the coefficient of determination ($R^2$ score):

\begin{equation} 
R^2 = 1-\frac{\sum_{i}{} (y_i - \hat{y_i})^2 }{\sum_{i}{} (y_i - \bar{y_i})^2}\,,
\label{eq: loss}
\end{equation}
where $y_i$ is the ground truth rating, $\hat{y_i}$ is the estimated rating, and $\bar{y_i}$ is the average of the ground truth rating. $R^2$ is a common metric to evaluate the fit of a regression prediction to the ground truth value. 






\section{Results \& Discussion}

\begin{figure*}[tb]
  \centering
  \begin{tabular}{@{}c@{}}
    \includegraphics[width=0.45\textwidth]{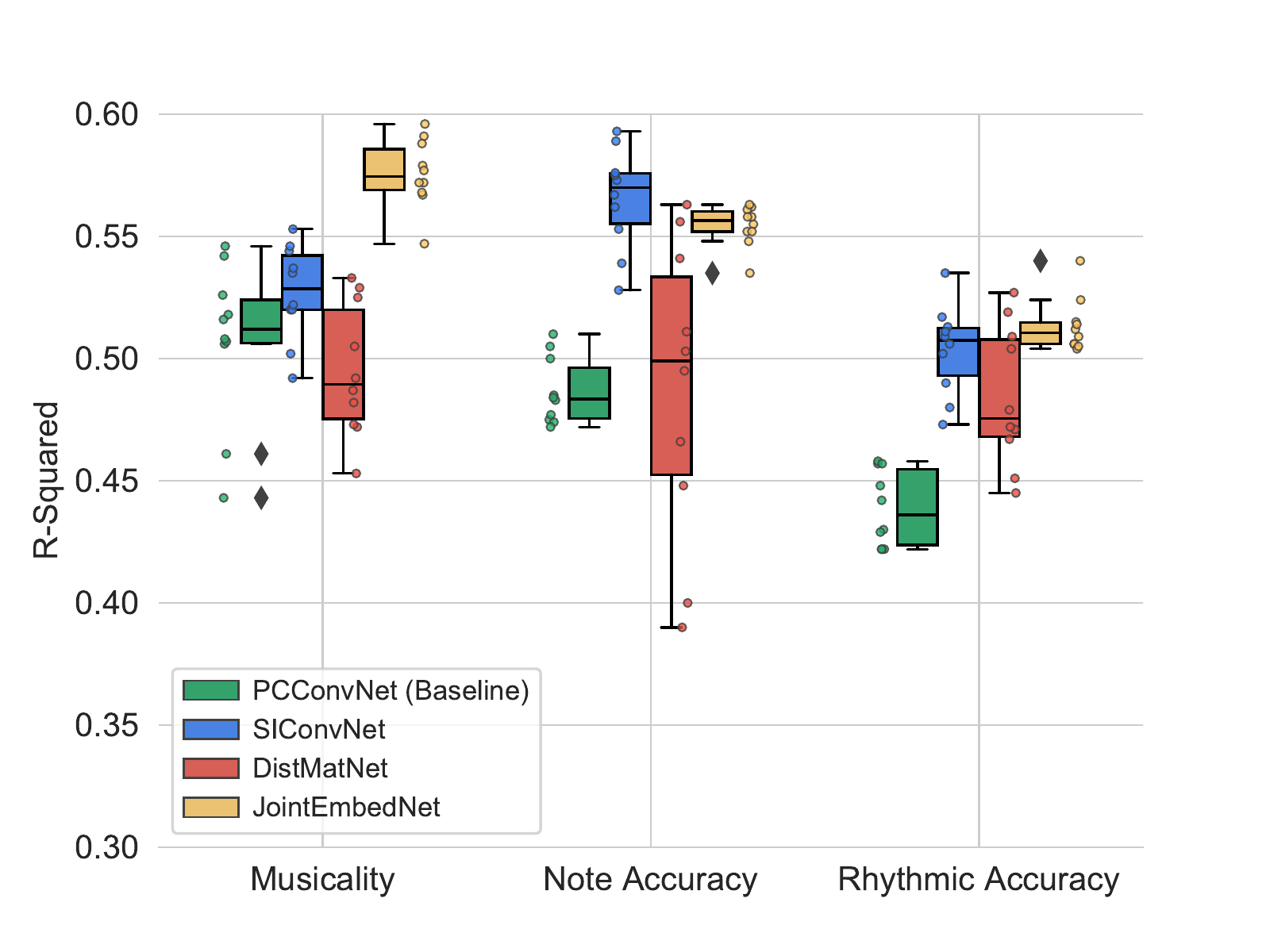} \\[\abovecaptionskip]
    \small (a) Middle School
  \end{tabular}
  \begin{tabular}{@{}c@{}}
    \includegraphics[width=0.45\textwidth]{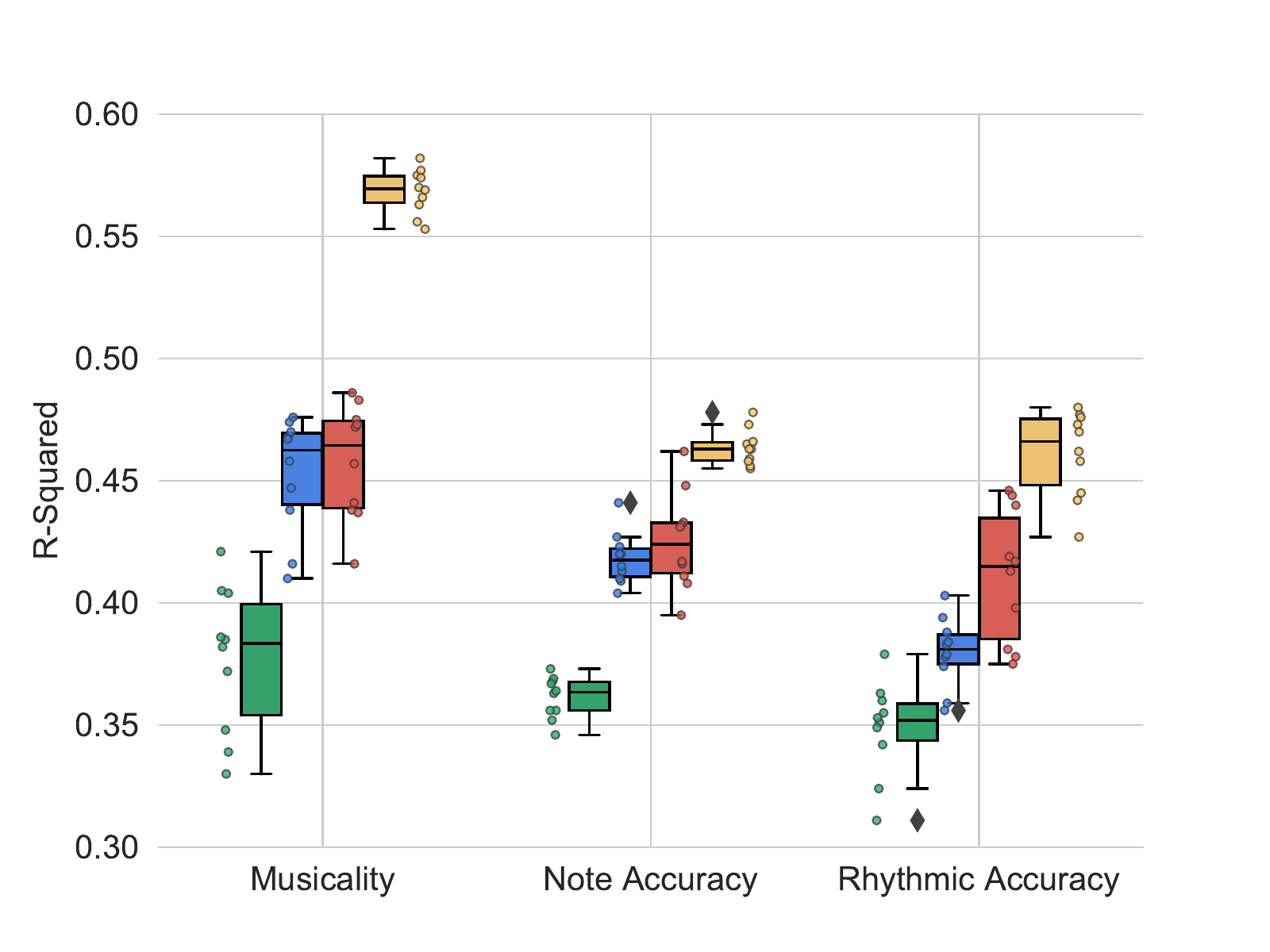} \\[\abovecaptionskip]
    \small (b) Symphonic Band
  \end{tabular}
  \caption{Box plots showing comparative performance (higher is better) across different models and assessment criteria.} 
  \label{fig:results_combined}
\end{figure*}

\subsection{Overall Performance}
\figref{fig:results_combined} shows the comparative performance for all models for middle school and symphonic band. We can make the following observations (\jiawen{with independent t-test results reported}):
\begin{compactenum}[(i)]
  \item We compare the performance of various models trained on different band performances. All systems perform better (higher $R^2$ value) on the middle school recordings than on the symphonic band recordings \jiawen{($p<0.01$ except JointEmbedNet for musicality)}. 
  One possible explanation for this is that symphonic band scores are usually more complicated and longer. For example, symphonic band scores tend to be performed at high tempo with high note density. The chunking into smaller lengths (and the downsampling of the distance matrix) compared to the score length might lead to a less accurate mapping to the assessment rating. An additional factor is that most performers in the symphonic band auditions exhibit greater skill level than middle school performers thus making it potentially more difficult to model the differences in proficiency levels.
  \item All score-informed models generally outperform the baseline, implying that score information is indeed helpful for MPA \jiawen{($p<0.01$ except SIConvNet for musicality, DistMatNet for musicality and note accuracy on middle school)}. 
  We notice, however, that the difference between the score-independent baseline and the score-informed models is smaller for the middle school than for symphonic band. Given the significant improvement over the baseline for symphonic band performances (which have complicated scores), we speculate that the score-informed models benefit more from access to score information. In other words, a score reference becomes more impactful with increasing proficiency level while the pitch contour alone contains most relevant information for medium proficiency levels.
  \item While the two models SIConvNet and JointEmbedNet both use the same input features, JointEmbedNet either outperforms or matches SIConvNet in all experiments.  The main difference between these two architectures is that SIConvNet simply performs a regression to estimate the assessments while JointEmbedNet learns a similarity in the embedding space to model the assessments. Therefore, we can assume that JointEmbedNet is able to explicitly model the differences between the input pitch contour and score especially in the case of symphonic band where the scores are more complicated.
  \item We observe that while DistMatNet and JointEmbedNet both utilize the similarity between the score and pitch contour, albeit at different stages of the network, JointEmbedNet typically performs better across categories and bands, and the gap \jiawen{is larger for musicality than for the other two categories}. It is possible that the absolute pitch at the input may be important for the final assessment (octave jumps, for example, would not be properly modeled in the distance matrix). More likely, however, is that the significantly larger input dimensionality of the matrix (compared to the aligned sequences for JointEmbedNet) negatively impacts performance. Most of the relevant information for MPA centers around the diagonal of the distance matrix with relatively small deviations depending on the students' tempo variation. Most of the distance matrix elements far from the diagonal contain redundant or irrelevant information, thus complicating the task. \jiawen{Another advantage that JointEmbedNet might have over DistMatNet in terms of overall assessment is that the distance is computed on the whole performance while DistMatNet computes a frame-level pitch distance, potentially complicating the task for overall quality measures like musicality.}
\end{compactenum}

\begin{figure}[tb]
  \centering
  \begin{tabular}{@{}c@{}}
    \includegraphics[width=0.45\textwidth]{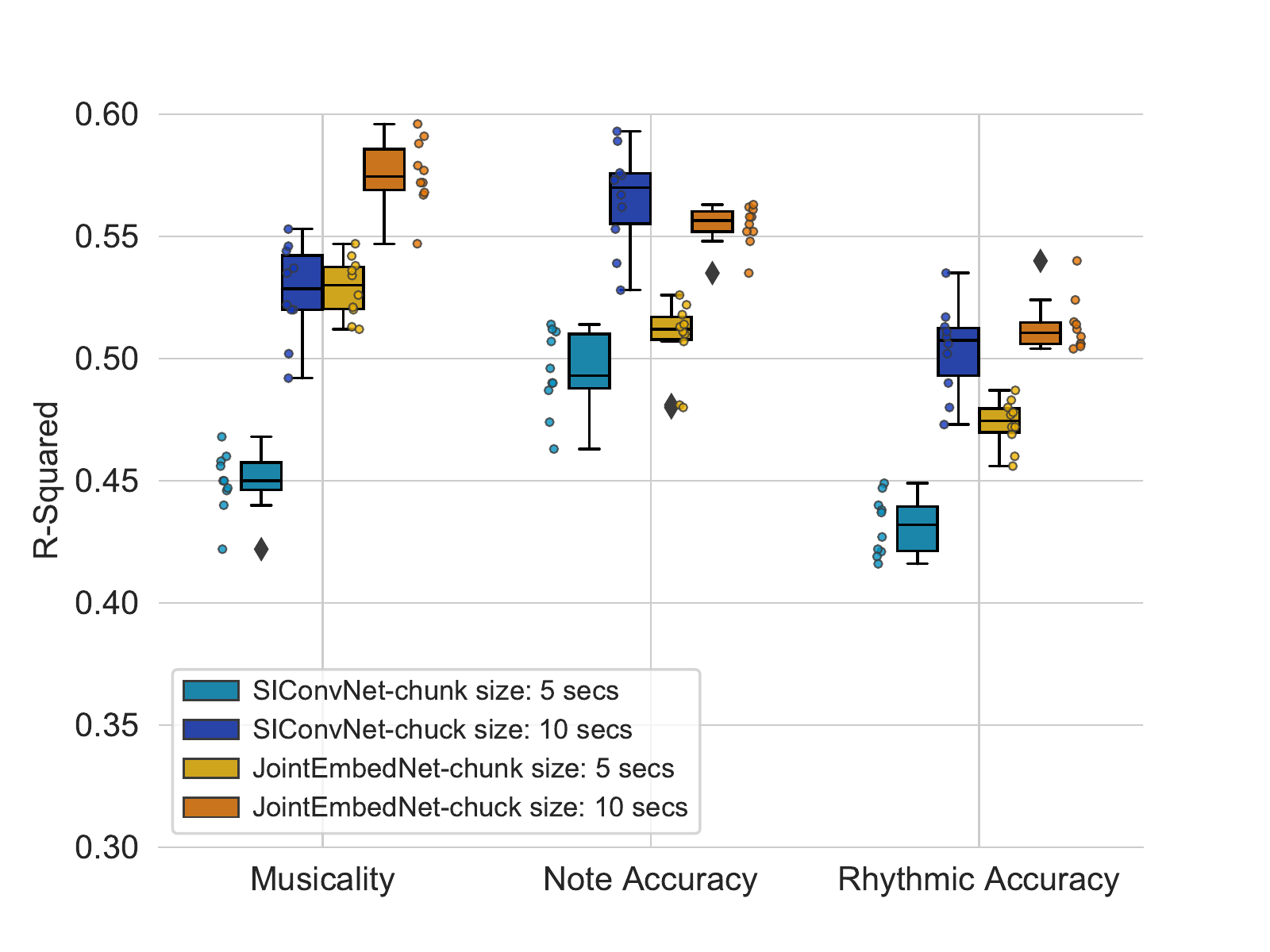} \\[\abovecaptionskip]
    \small (a) Middle School
  \end{tabular}
  \begin{tabular}{@{}c@{}}
    \includegraphics[width=0.45\textwidth]{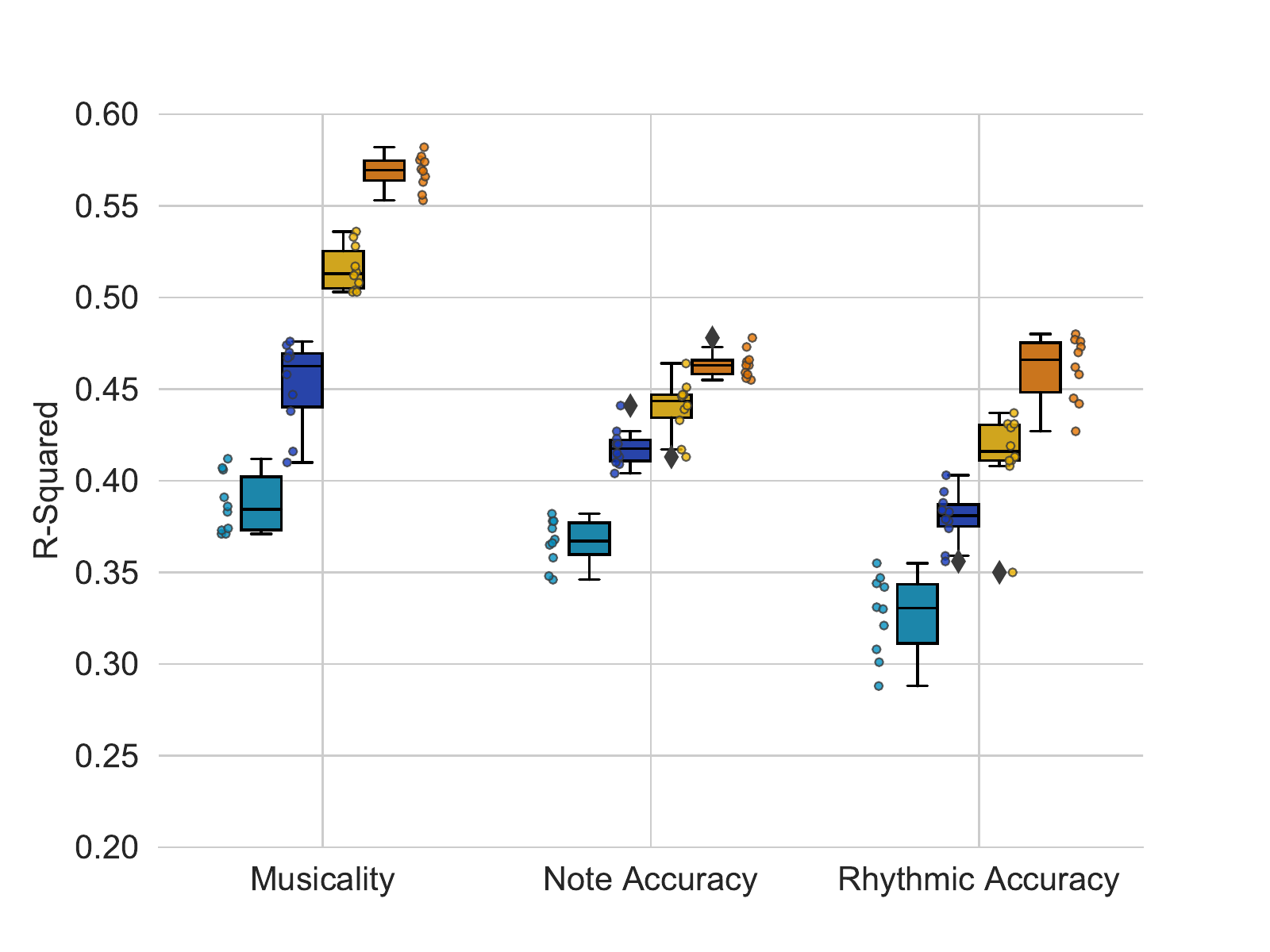} \\[\abovecaptionskip]
    \small (b) Symphonic Band
  \end{tabular}
  \caption{Box plots showing comparative performance (higher is better) across different chunk sizes for SIConvNet and JointEmbedNet.} 
  \label{fig:chunk_size_results}
\end{figure}

\subsection{Chunk Size}
In this experiment, we look at the impact of two different chunk sizes for the first two methods. \figref{fig:chunk_size_results} shows the results on middle school (a) and symphonic band (b). For both SIConvNet and JointEmbedNet, a chunk size of \unit[10]{s} outperforms that of \unit[5]{s} across all the bands.  

Chunking with random sampling is a form of data augmentation. By using the ground truth rating of the whole performance, the chunks are assumed to reflect the quality of the whole performance. The results show that \unit[5]{s} chunks might be too short to evaluate the whole performance while \unit[10]{s} chunks are much better suited regardless of category and score complexity. Chunk lengths greater than \unit[10]{s} were not tested because we restricted ourselves to the length of the shortest performance in the dataset. Consequently, we used a \unit[10]{s} chunk size for the experiment in \figref{fig:results_combined}.

\begin{figure}[tb]
  \centering
  \begin{tabular}{@{}c@{}}
    \includegraphics[width=0.45\textwidth]{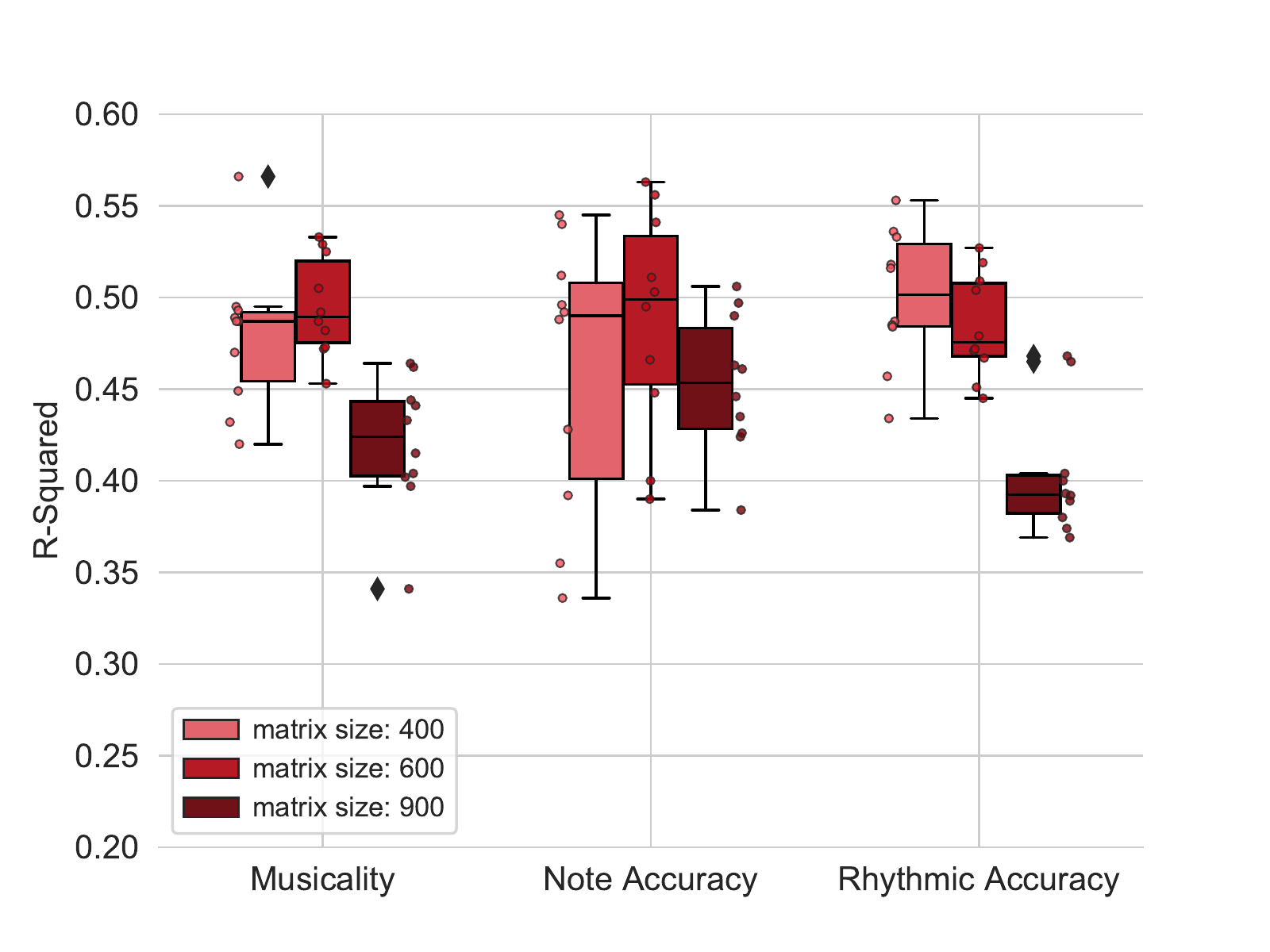} \\[\abovecaptionskip]
    \small (a) Middle School
  \end{tabular}
  \begin{tabular}{@{}c@{}}
    \includegraphics[width=0.45\textwidth]{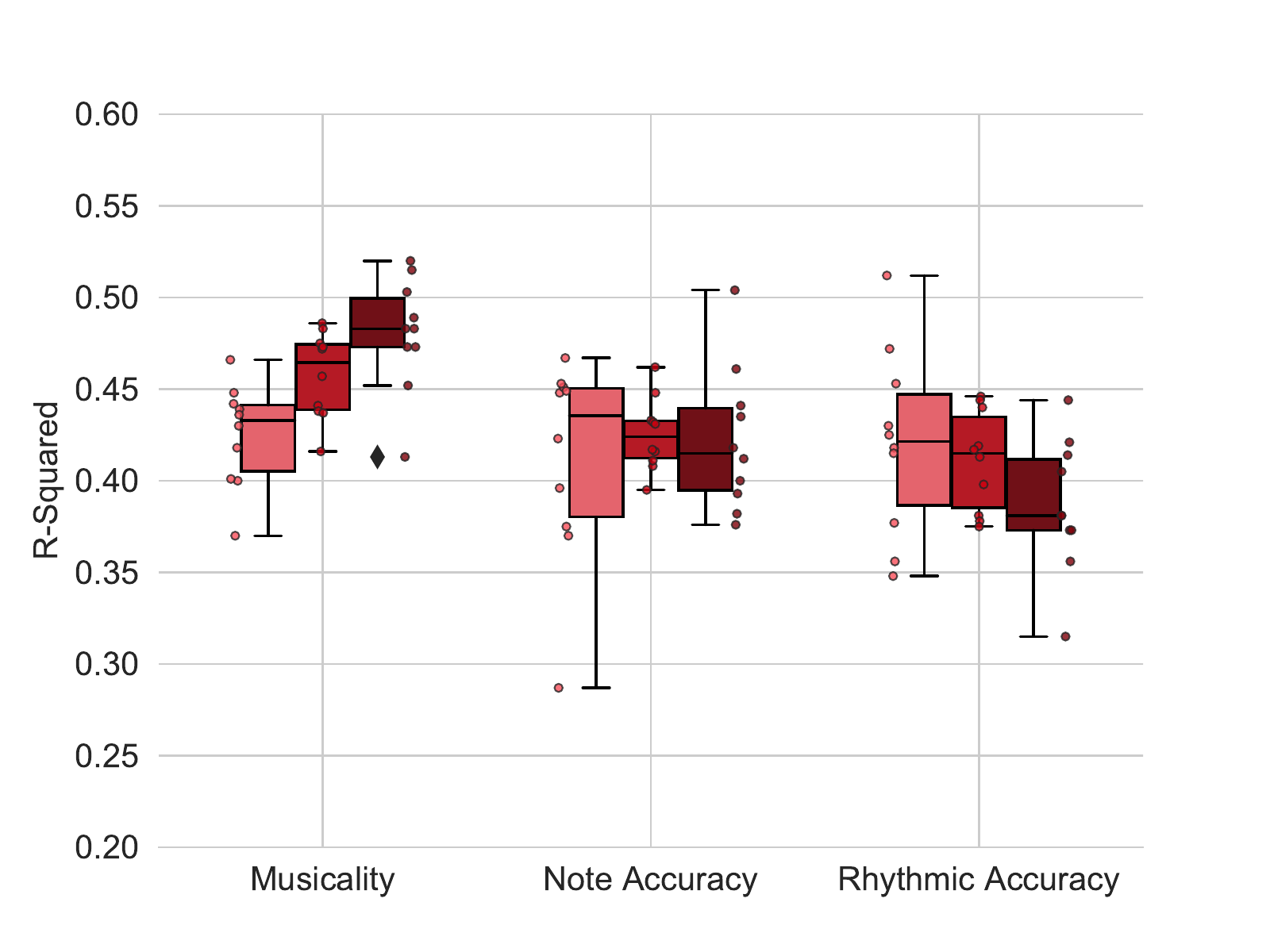} \\[\abovecaptionskip]
    \small (b) Symphonic Band
  \end{tabular}
  \caption{Box plots showing comparative performance (higher is better) across different matrix sizes for the distance matrix network (DistMatNet).} 
  \label{fig:dist_mat_results}
\end{figure}

\subsection{Distance Matrix Resolution}
In this experiment, we study the impact of the different input matrix resolutions $400\times 400$, $600\times 600$, and $900\times 900$, for the DistMatNet model. 
The results for both middle school and symphonic band are shown in \figref{fig:dist_mat_results}. 
First, the performance of rhythmic accuracy criterion tends to decrease with increasing distance matrix resolution. It might be more difficult for the same model structure to capture the complexity inside a larger matrix. This can also explain the result for middle school: although increasing the input resolution from $400\times 400$ to $600\times 600$ will capture more details, the performance decreases when the matrix resolution is further increased. 
Second, an input matrix size of $600\times 600$ leads to a slightly higher average score (0.46) on both symphonic and middle school than the other two resolutions (0.45 for $400\times 400$ and 0.43 for $900\times 900$). We ended up using the $600\times 600$ resolution for the experiment in \figref{fig:results_combined}.

\section{Conclusion}
This paper presents three novel neural network-based methods that combine score information with a pitch representation of an audio recording to assess a music performance. The methods include: 
\begin{inparaenum}[(i)]
    \item   a CNN with aligned pitch contour and score as the input, 
    \item   a joint embedding model that learns the assessment as the cosine similarity of the embeddings of both the aligned pitch contour and the score, and
    \item   a distance-matrix based CNN, using a differential representation of pitch contour and score at the input. 
\end{inparaenum}
The results show that all the methods outperform the score-independent baseline model. The joint embedding model achieves the highest average performance.

\ashis{Beyond the obvious applications in software-based music tutoring systems, score-informed performance assessment models (and objective MPA in general) can benefit the broader area of music performance analysis. Models capable of rating performances along different criteria could serve as useful tools for objective evaluation of generative systems of music performance. In addition, such models could also be explored for objective analysis of inter-annotator differences in rating music performances.}


In the future, we plan to incorporate timbre and dynamics information into the models as it has been shown to improve accuracy \cite{pati2018assessment}. This will also enable the model to assess performances in terms of tone quality, the criterion ignored in this study. We also plan to investigate other instruments and to examine cross-instrument relationships by training instrument-specific models. \jiawen{Furthermore, the musical score reference could be replaced with other representations such as the pitch contour of a highly-rated performance.} 

\section{Acknowledgment}\label{sec:acknowledgement}
We would like to thank the Florida Bandmasters Association for providing the dataset used in this study. We also gratefully acknowledge Microsoft Azure who supported this research by providing computing resources via the Microsoft Azure Sponsorship.

\bibliography{ISMIR2020template}

\end{document}